# Dijet Mass Resolution at the LHC

Dan Green
Fermilab

**Introduction**

The Large Hadron Collider (LHC) which will operate at CERN will provide unprecedented energies to be used in the search for new phenomena. However, it will also require that research be conducted at unprecedented luminosities. For example, at the LHC design luminosity, each event of interest will contain, on average, twenty minimum bias events. These events are embedded in an r.f. bunch crossing where individual events cannot be time resolved from the other events arising from the inelastic cross section. These "minbias" events will contribute to a "pileup" of energies when jets are being studied.

In order to reduce the influence of the pileup energy deposits in the calorimeters used at the LHC, techniques should be found to remove or reduce or correct for the pileup energies [1]. If this can be accomplished, then the resolution of important quantities for jet spectroscopy, such as the dijet mass may also be improved. In this note the use of auxiliary information within the event is examined.

**Jet Kinematics**

A jet is defined to be the final state remnant of a quark or gluon. If an ensemble of particles clustered in phase space is chosen, for example within an area of rapidity – azimuth space defined by a cone of radius Rc, then a jet axis is defined such that the total momentum of the ensemble perpendicular to that axis is zero [2]. With that choice of axis the jet four vector is the sum of the momentum and energy of the calorimeter clusters in the ensemble, which are assumed to be massless. Note, however, that the jet itself has a mass. Jet quantities in Eq.1 are uppercase; calorimeter clusters are lowercase.

$$\vec{P}_J = \sum_i \vec{p}_i, \quad E_J = \sum_i e_i \qquad (1)$$

$$\tan \phi_J = P_{yJ} / P_{xJ}$$
$$\tanh y_J = P_{zJ} / E_J$$

The dijet mass is then calculated using the four momenta of the two individual jets. A simplification occurs because the jet masses are small with respect to the dijet masses considered here, sequential Z bosons of mass 120 GeV. Therefore, for the purposes of dijet mass calculation the jet masses can be set to zero. Thus, to find the dijet mass, $M_{12}$, the individual jet masses, $M_1$ and $M_2$, are not required. Only three kinematic variables are needed for each jet, which we can take to be the transverse momentum, the pseudorapidity, and the azimuthal angle. The dijet mass depends only on these six variables which define the vector three momenta of the jets.

$$M_{12}^2 = 2E_{T1}E_{T2}[\cosh(y_1 - y_2) - \cos(\phi_1 - \phi_2)] \quad (2)$$
$$\sim 4P_{T1}P_{T2}\cosh^2(y_1 - y_2)/2$$

The dijets can approximately be taken to be back-to-back in azimuth for the Z(120) data set. Since the p-p C.M. frame is not the dijet C.M. frame, there is dependence of the dijet mass on the polar angles of the jets, as is evident in Eq.2.

There is additional information both within the jet phase space and outside it. The jet mass arises from the non-zero transverse momentum of the calorimeter clusters with respect to the jet axis. The squared jet mass can be written approximately as the sum of terms due to each calorimeter tower cluster.

$$M_J^2 = \sum_i m_i^2$$
$$\sim E_{TJ} \sum_i e_{Ti} r_i^2 \quad (3)$$
$$E_{TJ} \sim \sum_i e_{Ti} \cosh \delta\eta_i \sim \sum_i e_{Ti}$$

The jet mass represents additional information from within the jet cone. It depends on the details of the distribution of clusters with respect to the jet axis, as indicated by the dependence of the mass on $r_i = \sqrt{\delta\eta_i^2 + \delta\phi_i^2}$ where the quantities $\delta\eta_i$ and $\delta\phi_i$ refer to deviations of the cluster from the jet axis. The jet transverse momentum is approximately simply the scalar sum of the transverse momenta of the clusters.

Suppose a cluster from pileup energy deposits accidentally occurred in the cone of a jet. In that case the transverse momentum of the jet would increase, $P_{T1} \to P_{T1} + \Delta$. In turn, using Eq.2 and Eq.3, the mass of both the dijet and the jet itself would increase.

$$M_{12}^2 \to M_{12}^2[1+\Delta/P_{T1}]$$
$$M_1^2 \to M_1^2[1+\Delta/P_{T1}] + \Delta P_{T1} r_\Delta^2 \quad (4)$$
$$P_{T1} \to P_{T1} + \Delta$$

Clearly, it is expected that increases in dijet mass due to pileup will be correlated to increases in jet mass. Assuming that the intrinsic jet mass is small, the second term in Eq.4 for the change in jet mass is the most important one. Therefore, the correlation between jet and dijet mass is not simple because it depends on the deviation of the pileup cluster from the jet axis, $r_\Delta$. Multiple pileup clusters within the cone further complicate the situation.

There are other variables that can be defined within an event. For example for each jet the transverse momentum can be found in a cone of a given radius about an axis whose polar angle direction is defined by the rapidity of the jet but which is at 90 degrees in azimuth to the axis of the jet. This transverse momentum is presumably due to the underlying event and due to additional events which pile up on the bunch crossing of interest. It should serve to indicate the rough level of background jet activity in the particular event. These momenta are labeled as $P_{T1out}$, $P_{T2out}$ for the two jets in the dijet events.

**Z(120) Jets at LHC Design Luminosity**

A sample of events where a sequential gauge boson of 120 GeV mass, Z(120), was generated. They were then required to have both jets with transverse momenta larger than 20 GeV and pseudorapidities with magnitude less than 1.5. These events were passed through a full GEANT simulation of the CMS detector. At design luminosity, minimum bias events are overlapped with the events of interest, using a Poisson distribution of the number of overlap events. In this way a somewhat realistic model of events at LHC design luminosity, with time digitization and full modeling of the calorimetric energy resolution was performed [3].

The events were processed by using as input only the energy deposits in the towers of the electromagnetic and hadronic longitudinal compartments of the calorimetry. A simple cone algorithm was used with a seed tower. The transverse energy of the seed tower was required to be above 7 GeV. The mean number of seed towers found was 6.8. The mean transverse energy of the leading seed tower was found to be 31.6 GeV.

Using the seed as the initial axis of a cone of radius Rc, towers within that cone were added to the jet if they had a transverse momentum above a threshold of 0.5 GeV. After all towers were added, the axis of the jet was recalculated as defined by Eq.1. For a cone size of, Rc = 0.5, the leading jet in the event has a mean transverse momentum of 64.9 GeV. The average number of distinct clusters within the jet cone is 18.1 for the leading jet. Note that pileup is substantial. At 1/5 of LHC design luminosity, the average number of clusters is only 8.3, which indicates that about half of the accepted clusters are due to pileup.

The mean number of jets found by this simple procedure is 3.9 per LHC bunch crossing at the LHC design luminosity of $10^{34}/cm^2 \sec$. The calorimeter towers were assumed to be massless particles. Of necessity, the jets then have a finite mass. The mass of the leading jet was, on average $<M_{J1}> = 12.8$ GeV, while the next to leading jet has an average mass of $<M_{J2}> = 10.7$ GeV. The average value for the transverse momentum at 90 degrees in azimuth to the axis of the leading jet was, $<P_{T1out}> = 13.3$ GeV, compared to 64.9 GeV within the leading jet cone.

The mass of the dijet system is calculated by assuming that the jets are massless. The mean value of the dijet system mass is, $<M_{J1J2}> = 127.6$ GeV. The distribution of that mass is shown in Fig.1. Note that including the jet mass in the calculation of dijet mass does not improve the dijet mass fractional mass resolution nor does it shift the mean value substantially. In Fig.1 it is clear that there are events with rather large (> 300 GeV) dijet masses.

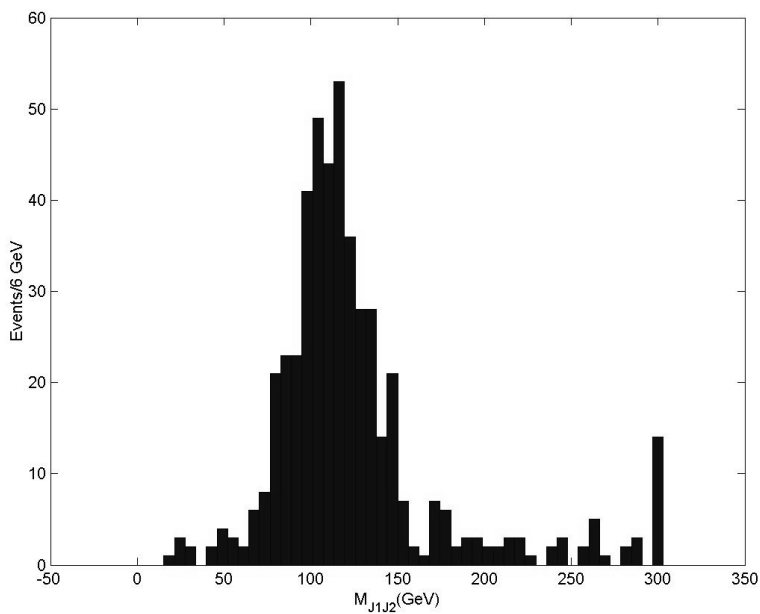

Figure 1: Dijet mass distribution for Z(120) events at LHC design luminosity with a jet cone of Rc=0.5. The jets are assumed to be massless.

**Jet Cone Radius**

The size of the cone containing the jet was roughly optimized for the LHC design luminosity. Three cone sizes were evaluated and the dijet mean mass and r.m.s. for this event sample were evaluated. The sample r.m.s. divided by the mean was found to be a minimum at a cone size of, Rc = 0.5. Some results are given in Table 1

## Table 1
## Jet Cone Radius Optimization

| Rc | $<n_i>$ | $<M_{J1}>$ GeV | $<M_{J1J2}>$ GeV | r.m.s./mean $M_{J1J2}$ |
|---|---|---|---|---|
| 0.3 | 9.6 | 10.4 | 101.5 | 0.72 |
| 0.5 | 18.1 | 12.8 | 127.6 | 0.61 |
| 0.7 | 29.3 | 15.5 | 184.9 | 0.89 |

Note that the number of towers grows rapidly with the cone radius because the phase space area is roughly uniformly populated and the area goes as the square of the radius. Therefore the area approximately doubles for the range of cones studied and the mean number of clusters also doubles.

Since the cluster tower contribution to the jet mass, Eq.4, is proportional to the radius from the jet axis, the tower contribution also increases as Rc increases, on average, roughly linearly with radius. The dijet mass almost doubles, on average, as the jet cone radius goes from 0.3 to 0.7. This increase is due partially to collecting more real jet fragments, but is also due to including background energy from underlying event and pileup events. At the LHC design luminosity, it is found that the sample r.m.s divided by the sample mean for the dijet mass has a minimum at Rc = 0.5 and this value is chosen for subsequent study.

It is clear that energy calibration of the calorimetry in LHC experiments will be dynamic. Looking at Table 1, it is observed that the mass scale depends approximately linearly on the jet cone radius. The mass scale also depends on the luminosity.

**Jet Mass and a Cut on Clusters**

The mass contribution of an individual tower, $m_i$, folded in quadrature in the overall jet mass sum, is approximately;

$$m_i \sim r_i \sqrt{e_{Ti} E_{TJ}} \qquad (5)$$

The mean value of a calorimeter tower contribution to the jet mass is 2.1 GeV.

As seen in Eq.3 and Eq.5, pileup will contribute to the jet mass and the dijet mass. In Fig.2 is shown the individual contribution of the tower cluster to the jet mass. Individual calorimeter clusters contributing ~ 10 GeV to the mass are observed.

Clusters were removed from the jet sum if they occurred with radius > 0.1 and if they contributed more than 9 GeV to the jet mass. Harder cuts were attempted but were found to compromise the dijet mass resolution. Only a few clusters were removed from each jet, about 0.4 clusters on average. Nevertheless, events with very large dijet masses had their mass reduced considerably.

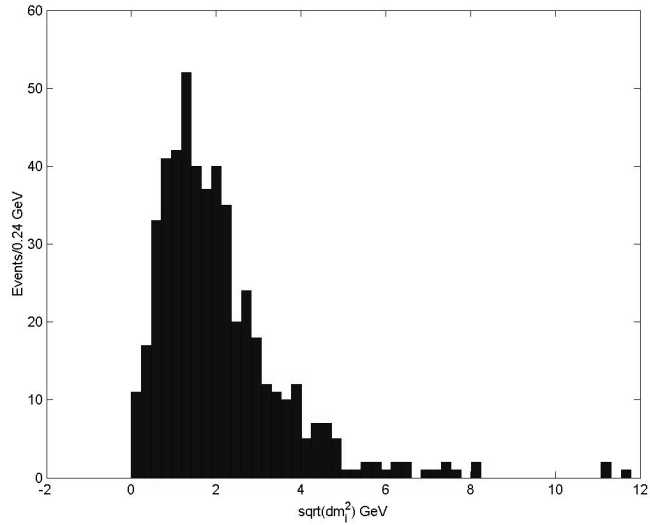

Figure 2: Distribution of the mass contribution, $m_i$, to the jet mass made by an individual calorimeter cluster.

The resulting mass distribution after the cluster mass cut is shown in Fig.3. A Gaussian fit was made to the mass distribution over the range (80,140) GeV in 6 GeV mass bins. The fitted mass was 111.0 GeV with a standard deviation of 19.8 GeV (17.9 %). The chi-square per degree of freedom was 10.75/7, indicating an adequate fit.

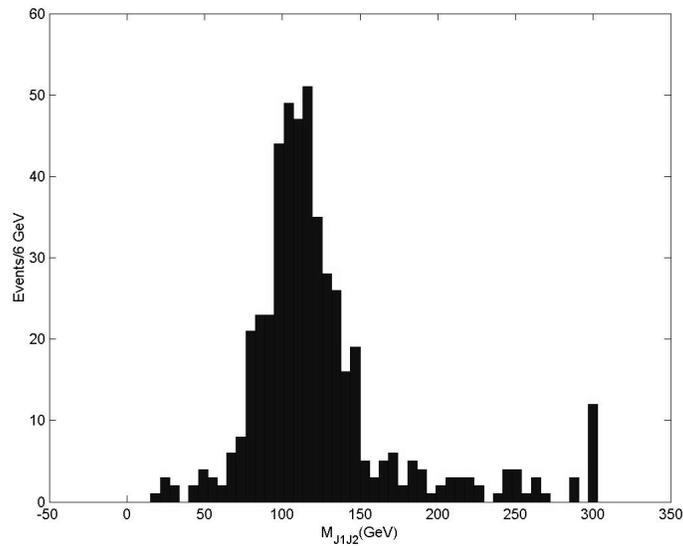

Figure 3: Mass distribution in Z(120) LHC events after removing calorimeter clusters with r > 0.1 and a mass term > 9 GeV.

**Jet Mass and a Correction for Dijet Mass**

Clearly, the parton that fragments into the jet has a small mass after having shed it's virtuality in a partonic shower. Equally clearly, in the presence of pileup and transverse momentum from the jet fragmentation process the jets will appear to have a mass, one which varies with luminosity and with parameters such as cone size. The mass will then fluctuate on an event-by-event basis. The observed mass distribution of the leading jet in the present Z(120) sample is shown in Fig.4. The mean jet mass is 12.8 GeV and the distribution in mass is rather wide. Very large jet masses have already been eliminated by the 9 GeV cluster mass cut described above.

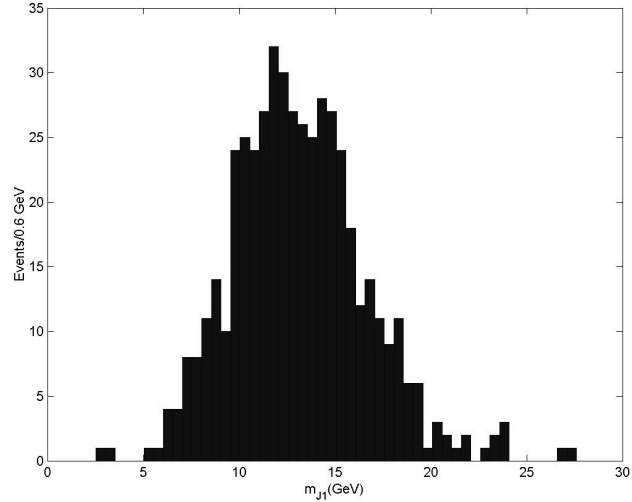

Figure 4: Distribution of the mass of the leading jet in Z(120) events at the LHC for a cone size of Rc = 0.5 and LHC design luminosity.

In Fig.5 the distribution of transverse momenta within a cone of radius Rc = 0.5 at 90 degrees in azimuth to the leading jet is shown. This quantity sets the scale within the event for the momenta of pileup events, on average, and the average of the underlying event.

If the large variations in the masses of the dijets, the jets and $P_{Tout}$ are ascribed to large pileup fluctuations in the bunch crossings, then there should be an event by event correlation between the dijet mass and the individual jet masses. Based on Eq.4, if there is only a single pileup cluster in only one of the jets, a very approximate corrected dijet mass is given in Eq.6.

$$M_{12} \sim M'_{12} / \sqrt{1 + (M_1 / r_\Delta (P_{T1} - <P_{T1out}>))^2 + ...} \qquad (6)$$

In Fig.6 the mass of the dijet is plotted vs. the event-by-event sum of the jet masses with the average transverse momentum outside the jets at 90 degrees in azimuth subtracted. This quantity, although motivated by Eq.4, is purely an experimental attempt to improve the dijet mass resolution at high LHC luminosity. The simplest linear form was adopted in this initial study, whereas from Eq.6 it might be expected that the correction term is roughly, $[M_1 / r_\Delta (P_{T1} - P_{T1out})]^2$ with another term corresponding to pileup contributions from the second jet.

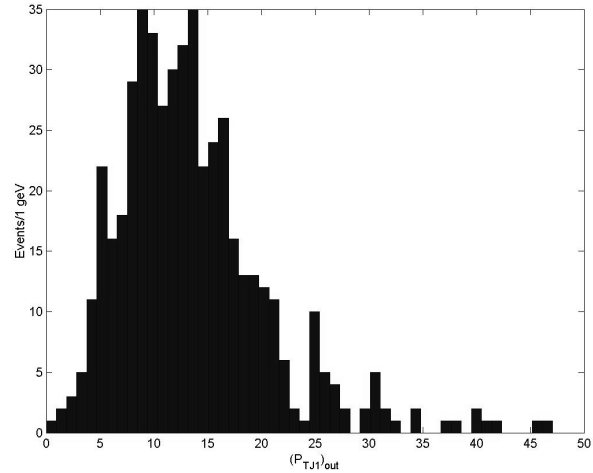

Figure 5: Distribution of the transverse momentum in a cone of radius Rc = 0.5 with the cone axis defined to have the rapidity of the leading jet but to have an azimuthal angle rotated 90 degrees to the leading jet.

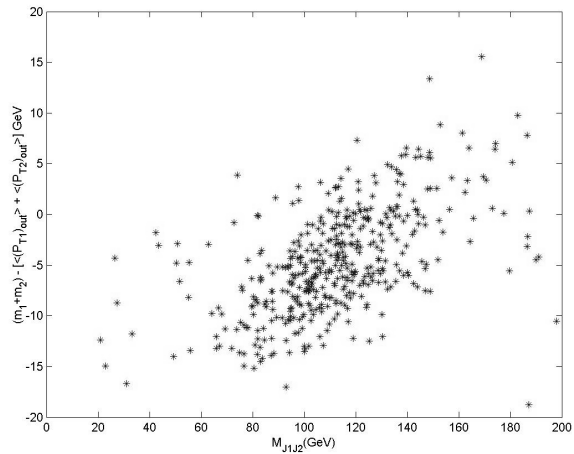

Figure 6: Scatter plot of the dijet mass vs. the sum of the masses of the two jets making up the dijet. The mean transverse momentum in the event sample within cones of size Rc = 0.5 oriented at 90 degrees in azimuth to the jets is subtracted from the jet mass.

Clearly there is a correlation between large jet mass and large dijet mass which appears to indicate bunch crossings with large pileup fluctuations. The correlation that is observed in Fig.6 is used to linearly correct the value of the dijet mass. The assumption made is, fundamentally, that the jet mass should, physically, be small. The average out of cone transverse momentum for the two cones is subtracted from the jet mass. The dijet mass is calculated assuming zero jet mass, and any deviation of the jet mass is put in as a correction to the dijet mass. Therefore, the mean dijet mass is not much altered by this correction for this particular event sample. The parameter a in Eq.7 is determined by examining the mass distributions and choosing value yielding the minimal dijet mass resolution, a = 2.7.

$$M'_{12} = M_{12} - a[(M_1 - <P_{T1out}>) + (M_2 - <P_{T2out}>)] \qquad (7)$$

The corrected dijet mass distribution is shown in Fig.7. Clearly, the mass resolution has been improved by making use of this correction. In addition, the events observed in Fig.3 which remained with a large dijet mass after the cluster mass cut have been much reduced. This correction then appears to alleviate some of the worst effects of large pileup fluctuations.

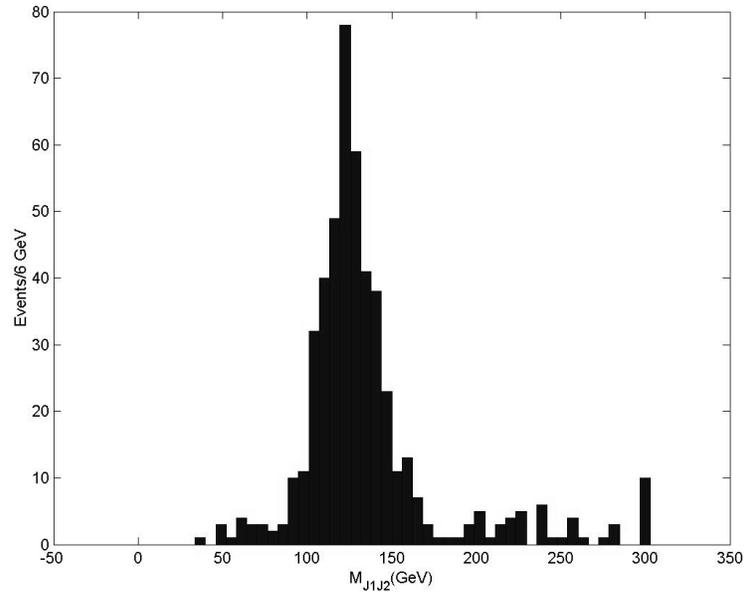

Figure 7: Dijet mass distribution for Z(120) events at the LHC. A correction has been applied in calculating the dijet mass using the event by event measurements of the masses of the two jets which make up the dijet system and average values of the out of cone transverse momentum. The horizontal scale is the same as is used in Fig.1 and Fig. 3 for comparison.

The corrected dijet mass distribution has been fit in the mass range (95,155) GeV in 6 GeV bins. The fit results are displayed in Fig.8. The fitted mean is 124.3 GeV with a standard deviation of 14.93 GeV (12.0 %). The chi-square per degree of freedom is 7.8/7 which indicates that after correction the shape is still quite Gaussian. The quality of the fit is evident in Fig.8.

The fitted standard deviation divided by the mean, or percent mass resolution, is 17.9 % for the uncorrected dijet mass (Fig. 3) and 12.0 % for the corrected mass. This is a roughly 50% improvement in mass resolution. This improvement justifies a more detailed future study. For example, different LHC luminosities should be explored and resonances at different masses. Given the ad hoc nature of the correction applied here, there is no guarantee that the procedure will work for different event samples. Also, one expects that this improvement should factorize from improvements in the measurement of the cluster energy that are provided by "energy flow" algorithms [4].

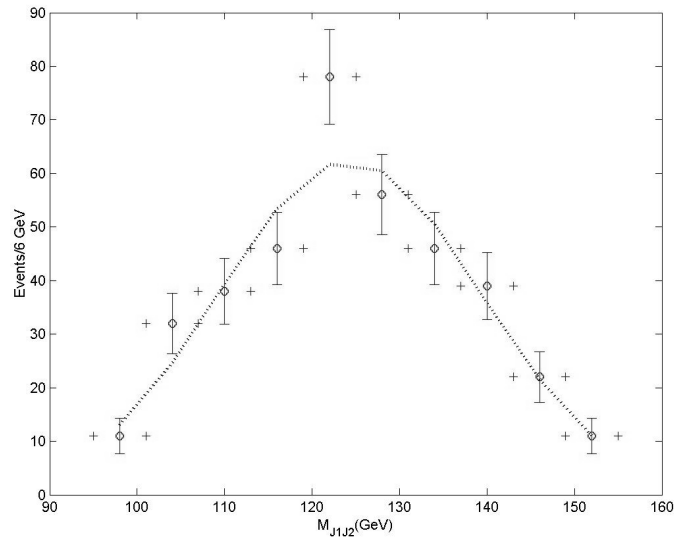

Figure 8: Mass distribution and Gaussian fit for the corrected dijet mass distribution in Z(120) LHC events shown in Fig.7.

**Summary**

The mass resolution for a sample of Z(120) narrow resonance events produced at the LHC at design luminosity with a realistic detector simulation has been studied. Using additional kinematic variables beyond those required to compute the dijet mass resolution, a correction factor was devised. The variables used both out of cone and in cone energy deposit. Applying this factor a 50% improvement in the mass resolution for the particular event sample was obtained. Heartened by this progress, further studies will be undertaken in the future.

**References**

1. A. Beretvas et al., Fermilab – FN – 0626 (1994)

2. A. Baden, Int. J. of Mod. Phy. A, (1997)
3. S. Kunori – private communication.
4. For example, S. Magill in "Calorimetry in Particle Physics, Ed. R-Y Zhu, World Scientific (2002) or D. Green, Fermilab-FN-0709 (2001)